# Ultra-compact and broadband tunable mid-infrared multimode interference splitter based on graphene plasmonic waveguide

Ruiqi Zheng,[1] Dingshan Gao, [1,2*] and Jianji Dong[1]

[1]*Wuhan National Laboratory for Optoelectronics, Huazhong University of Science and Technology, Wuhan 430074, China*

[2]*State Key Laboratory on Integrated Optoelectronics, College of Electronic Science and Engineering, Jilin University, Changchun 130012, People's Republic of China*

*\*Corresponding author: dsgao@hust.edu.cn*

We propose and design an ultra-compact and broadband tunable multimode interference (MMI) splitter in mid-infrared based on graphene plasmonic waveguides. The size of the device is only 0.56μm×1.2μm, which corresponds to device area of only about $0.014\lambda^2$, where λ is the vacuum wavelength. And the center wavelength of the device can be tuned in a broad band from 7μm to 9μm with the Fermi level of graphene varied from 0.5eV to 1eV. Furthermore, the device is easy to be fabricated on chip.

Surface plasmon polaritons (SPP) waveguides can provide sub-wavelength confinement beyond the diffraction limit and increase the integration density of the photonic circuit [1][2]. Mid-infrared SPP waveguides attract increasing interests for chemical sensing since many molecules such as methane and carbon dioxide exhibit fundamental vibrational absorption resonances in the mid-infrared wavelength range [3][4]. However, due to the huge negative permittivity of noble metal in mid-infrared, the normal metal/dielectric SPP waveguides have very poor confinement, which constrains the device sizes and integration density [5]. As a candidate, graphene, a novel two-dimensional material of carbon atoms arranged in honeycomb lattice, supports extremely confined graphene surface plasmons (GSPs) in mi-infrared and THz spectral [6][7]. Besides the strong confinement of GSPs modes, the electrical and optical properties of graphene can be modified by tuning Fermi level EF of graphene, via chemical doping or electrical gating [7]. Different kinds of GSPs waveguides for mid-infrared have been studied, such as edge GSP (EGSP)[8], waveguide GSP (WGSP)[8], dielectric loaded graphene plasmonic waveguide (DLGPW) [9] and so on [10-12].

The multimode interference (MMI) splitter [13] is an important component to realize many functions, such as beam splitter [14], add-drop multiplexing [15], ratio controllable splitter [16], and so on [17-19]. Recently, a mid-infrared MMI splitter was reported based on Germanium-on-Silicon (GeOS) rib waveguide [20]. However, the length of the MMI splitter is relatively long, which is about 60μm. Moreover, the bending radius of GeOS rib waveguides is as large as 350μm. So GeOS platform is not suitable for the further high-density integration. And previously reported MMI splitters can only work at a definite center wavelength or within a narrow bandwidth. So for the wideband applications, repeated designs for different working wavelengths are necessary. And to our knowledge, there is not any mid-infrared MMI splitter based on SPP waveguides reported so far.

In this letter, we propose an ultra-compact and broadband tunable MMI 3dB splitter based on GSP waveguides in mid-infrared. The length of the splitter is only about 1.2μm and the working wavelength can be freely tuned from 7μm to 9μm by shifting the Fermi level of graphene from 0.5eV to 1eV. Besides, the splitter can easily be fabricated on chip.

The schematic of our MMI splitter is shown in Fig. 1(a). A thin graphene layer with thickness t = 0.5nm is sandwiched between the dielectric substrate and the above dielectric strip. The graphene and the button metal layer act as the electrodes for applying voltage. The MMI structure is formed on the above dielectric strip. The relative permittivity of the dielectric strip and the substrate is 3.92, and their thicknesses are $h_1$ = 0.1μm, $h_2$ = 0.1μm respectively. The input and output waveguides of the MMI splitter are both single mode waveguides, which cross-section (xy-plane) is shown in the inset of Fig. 1(a). The top view of the whole MMI splitter is shown in Fig. 1(b), with the size parameters labeled on it.

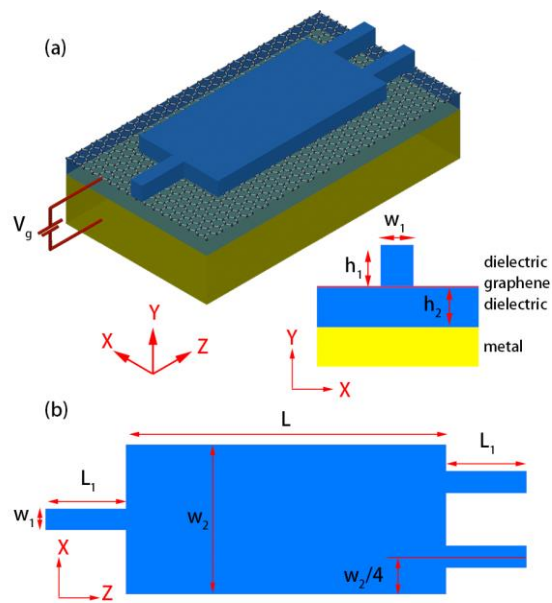

**Fig. 1.** (a) Schematic of the tunable MMI 3dB splitter, where the above and below blue layers are the MMI structure and the dielectric substrate respectively, which are separated by a thin graphene layer. The button yellow layer is the metal electrode. The voltage is added on the graphene and the metal electrode. The inset shows the cross-section (xy-plane) of the input/output single mode waveguide. (b) The top view of the whole device (xz-plane) with the size parameters labeled on it.

According to the self-imaging theory of MMI [13], the even and odd modes of the multimode waveguide should have cosine and sine shape electric field profiles respectively. However, in the

EGSP or WGSP, the electrical fields of the modes mainly concentrate on the edges of the waveguides [8], which will eliminate the self-imaging effect at all. The DLGPW has perfect cosine or sine shape mode profile, so it is a better choice for the MMI device. Moreover, without the difficulty to control the edge shape with desired atomic arrangements in EGSP and WGSP, DLGPW can be fabricated easily on a whole layer of graphene [9].

To design the MMI splitter, the modes of the DLGPW need to be calculated to determine suitable widths of the single mode and multimode waveguides. So we employ the finite element method-based (FEM) software (COMSOL Multiphysics) to solve the modes of the DLGPW. During the simulation we set the Fermi level of graphene $E_F$ = 0.8eV and the wavelength $\lambda$ = 8μm. And the graphene is treated as an anisotropic media [9]. The in-plane relative permittivity of the graphene $\varepsilon_\parallel = 2.5 + i\sigma_g/\omega\varepsilon_0 t$, where, $\varepsilon_0$ is the vacuum permittivity, $\omega$ is the angular frequency and $\sigma_g$ is the optical conductivity of graphene. The out-plane relative permittivity of the graphene $e_\wedge = 1$. Here, the conductivity of graphene is from the Kubo formula [5] such as

$$\sigma_{intra}(\omega) = \frac{2e^2 k_B T}{\pi \hbar} \frac{i}{\omega + i\tau^{-1}} \ln(2\cosh\frac{E_F}{2k_B T})$$

$$\sigma_{inter}(\omega) = \frac{e^2}{4\hbar^2}[H(\frac{\omega}{2}) + i\frac{4\omega}{\pi}\int_0^\infty \frac{H(\theta) - H(\frac{\omega}{2})}{\omega^2 - 4\theta^2}d\theta]$$

$$\sigma_g(\omega) = \sigma_{intra}(\omega) + \sigma_{inter}(\omega) \quad (1)$$

$$H(\theta) = \frac{\sinh\frac{\hbar\omega}{k_B T}}{\cosh\frac{E_F}{k_B T} + \cosh\frac{\hbar\omega}{k_B T}}$$

Where, $\sigma_{intra}$ and $\sigma_{inter}$ are the contributions from intraband transitions and interband transitions respectively. T = 300K, $V_F = 10^6$ m/s, $\tau = \mu E_F/eV_F^2$, $\hbar$ is the reduced Planck constant, $k_B$ is the Boltzmann constant, $\omega$ is the angular frequency, $V_F$ is the Fermi velocity, $\mu$ is the carrier mobility of graphene. The carrier mobility of graphene ranges from $\mu$ = 1000 cm$^2$/(V·s), in which grows via chemical vapor deposition (CVD)[21], to $\mu$ = 230000 cm$^2$/(V·s) in suspended exfoliated graphene [22]. Here we choose $\mu$ = 10000 cm$^2$/(V·s), $E_F$ = 0.8eV and $\tau \approx$ 0.5ps.

The simulated electric field profile of the fundamental mode of the DLGPW (w = 0.32μm) is shown in Fig. 2(a). The white line gives the lateral electric field profile of the fundamental mode along x direction, which is similar to cosine function. So the DLGPW is suitable for the MMI splitter. Besides, the mode profiles of the first 4 lowest order modes of the DLGPW (w = 0.32μm) are shown in Fig. 2(b), which also have good cosine or sine distributions along x direction.

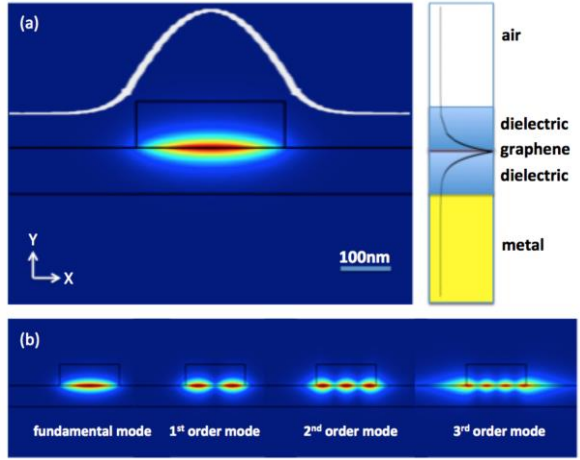

**Fig. 2.** (a) The electric field distribution of the fundamental mode of the DLGPW (xy-plane) with w = 0.32μm, $E_F$ = 0.8eV, λ = 8μm, $h_1$ = 0.1μm and $h_2$ = 0.1μm. The white line is the electric field profile of the fundamental mode along x direction, which is similar to the cosine function. The right diagram shows the vertical electric field distribution of the fundamental mode. (b) Mode profiles of the first 4 lowest order modes of the DLGPW with w = 0.32μm.

To determine the appropriate widths for the single mode and multimode waveguides, we also calculated the mode effective indexes of the DLGPW under different dielectric strip widths, as shown in Fig. 3.

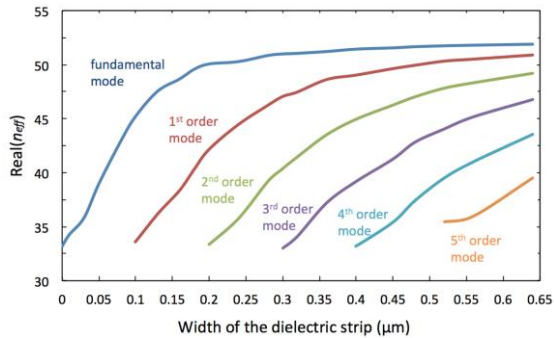

**Fig. 3.** The relationship between the real part of the mode effective indexes of different modes of the DLGPW Real(neff) and the width of the dielectric strip.

To make trade-off between the self-imaging quality and the length of the device, we choose $w_1$ = 80nm and $w_2$ = 560nm. According to the self-imaging theory of MMI, the intermodal beat length between the lowest two order modes $L_\pi = \pi/(\beta_0-\beta_1) = \lambda/[2(n_0-n_1)]$, where $n_0$, $n_1$ are the real part of the effective index of fundamental mode and first order mode respectively. The self-imaging distance is $L_{1f} = 3L_\pi/4$, and the two-fold image distance is $L_{2f} = L_{1f}/2 = 3L_\pi/8$. Due to the large effective index contrast (Δn = $n_0-n_1$ = 1.286) of the fundamental mode and the first order mode of the DLGPW, the two-fold image distance of the MMI is very short as $L_{2f}$ = 1.166μm, which is only about λ/7.

Besides the ultra-compact size, our splitter can be freely tuned in a broad bandwidth from 7μm to 9μm with the Fermi level of the graphene varied from 0.5eV to 1eV. Looking back to the Kubo formula in Eq (1), the conductivity of graphene σg is dependent on the wavelength λ and the Fermi level $E_F$. As a result, with a fixed length of the MMI region (L = $L_{2f}$ = 1.2μm), we can tune the Fermi level of the graphene by changing the voltage to shift the working wavelength of our MMI splitter.

Therefore we use the 3-D finite-difference time-domain (3D-FDTD) method to simulate the light propagation in our MMI splitter. The width of the single mode input/output waveguide is $w_1$ = 0.08μm and that of the multimode waveguide is $w_2$ = 0.56μm. The length of the MMI is set long enough to get two-fold images. At first, the light propagation in the MMI was simulated for $E_F$ = 0.8eV and λ= 8μm, as shown in Fig. 4. From Fig. 4(a), we can find very clear self-imaging phenomenon in the MMI. Fig. 4(b) and (c) are the cross section field profiles of the input fundamental mode and the two-fold images. Then from Fig. 4(a) we can also get the accurate two-fold imaging distance $L_{2f}$ = 1.12μm, where the white dashed line locates. This value is very close to the analytic calculation result $L_{2f}$ = 1.166μm from self-imaging theory. So the analytical calculation and numerical simulation prove that our MMI splitter has an ultra-compact size.

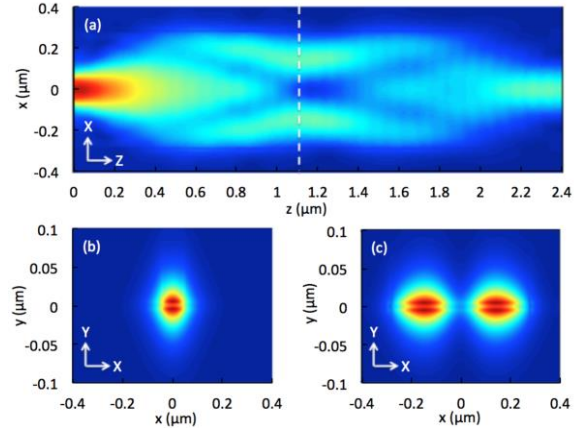

**Fig. 4.** (a) The top view of the electric field profile in the MMI splitter, with $w_2$ = 0.56μm, $E_F$ = 0.8eV, λ = 8μm. The two-fold image distance of the MMI region is $L_{2f}$ = 1.12μm, labeled by the white dashed line. (b) Cross section electric field profile of the single mode input waveguide with width of $w_1$ = 0.08μm. (c) Cross section electric field profile of the two-fold images at z = 1.12μm.

Further, to examine the tunability of our MMI splitter, we fix the two-fold imaging distance as $L_{2f}$ = 1.2μm and get the dependence between the working wavelength λ and the Fermi level of the graphene $E_F$. With the previously mentioned simulation method, we can get a 2D colored map (Fig. 5) to show the two-fold imaging distance $L_{2f}$ changing with the Fermi level $E_F$ of the graphene and the working wavelength λ. The figure is got by 2-D interpolation after sweeping the Fermi levels and the working wavelengths. And the red line is the fitting line representing the relationship between Fermi levels $E_F$ and the center wavelengths $\lambda$ when $L_{2f}$ = 1.2μm. The red line shows that the working wavelength for MMI splitter with L = 1.2μm is almost linearly varied with the Fermi level.

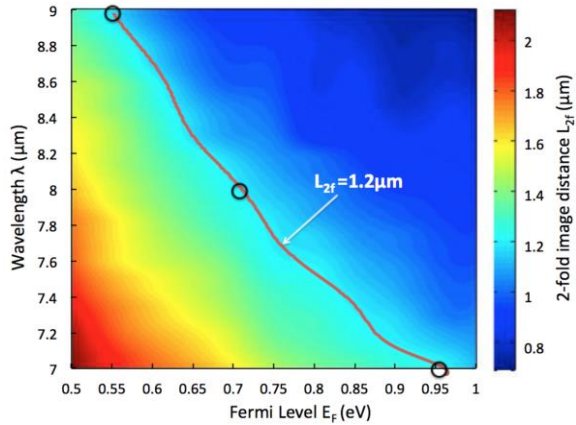

**Fig. 5.** The two-fold image distance of MMI $L_{2f}$ varying with the Fermi level of the graphene $E_F$ and the wavelength $\lambda$. The value of $L_{2f}$ is represented by the color. The red line gives the relationship between Fermi levels and wavelengths when $L_{2f}$ = 1.2μm. And the three point labeled with black circles on the red line are selected to test the MMI splitting capability.

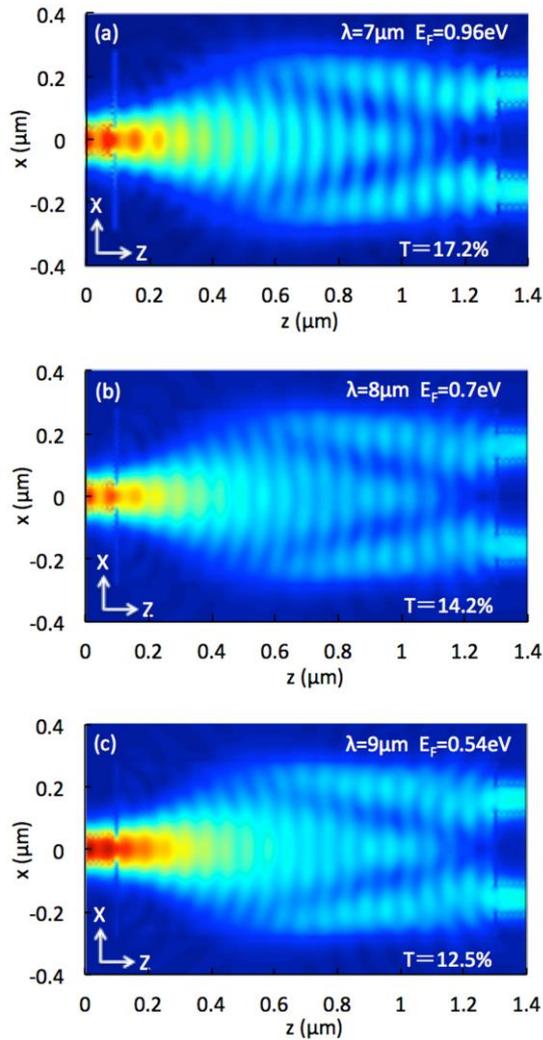

**Fig. 6.** The simulated electric field profiles for the three points in Fig. 5. Here T is the transmittance of each output waveguide. (a) $E_F$ = 0.96eV, λ= 7μm (b) $E_F$ = 0.7eV, λ = 8μm (c) $E_F$ = 0.54eV, λ = 9μm.

To further demonstrate the tunability of our MMI splitter, three points labeled with black circles on the red line in Fig. 5 are selected to make 3D-FDTD simulations. Their working wavelengths λ are 7μm, 8μm and 9μm, and corresponding Fermi levels $E_F$ of graphene are 0.96eV, 0.7eV and 0.54eV. The simulated electric field distributions of the MMI splitter for these three points are shown in Fig. 6(a)-(c). We can see that, the imaging qualities and splitting behaviors are very good for all three wavelengths, even though the length of MMI is fixed as 1.2μm. So our MMI splitter is not only ultra-compact in size but also tunable in a wide band of 7μm~9μm.

In summary, we have proposed and designed an ultra-compact tunable MMI 3dB splitter in mid-infrared based on the dielectric loaded graphene plasmonic waveguide. The working wavelength of the device can be electrically tuned from 7μm to 9μm with the Fermi level varied from 0.5eV to 1eV. The tunability of our device eliminates the complex device redesign or refabricating processes for the broadband applications. Moreover, the footprint of the whole device is only 0.56μm×1.2μm, which the size is only about $0.014\lambda^2$ in mid-infrared. Moreover, the device can be easily fabricated on a whole graphene layer without need to etch the graphene. The proposed ultra-compact and electrically broadband tunable MMI splitter can be used to form more complex mid-infrared photonic circuits on chip, such as the Mach-Zehnder interferometer (MZI) or optical switch based on MMI, etc.

This work was partially supported by the National Natural Science Foundation of China (Grant No. 11374115 and 61261130586), the Fundamental Research Funds for the Central Universities of China (Grant No. 2015TS080), and the Opened Fund of the State Key Laboratory on Integrated Optoelectronics, China (Grant No. 2011KFJ002).